\documentclass[aps,pra,prabib,showpacs,superscriptaddress]{revtex4}
\usepackage{psfig}
\usepackage{graphicx}
\usepackage{amsmath}
\usepackage{amssymb}
\usepackage{amsfonts}
\usepackage{bm}

\begin{document}

\title{The non-classical scissors mode of a vortex lattice in a Bose-Einstein condensate}

\affiliation{Van der Waals-Zeeman Instituut, Universiteit van Amsterdam, 
Valckenierstraat 65, 1018 XE Amsterdam, The Netherlands}
\affiliation{Laboratoire Kastler Brossel, \'Ecole Normale
Sup\'erieure, 24 rue Lhomond, 75231 Paris Cedex 05, France}

\author{Carlos Lobo}
\affiliation{Van der Waals-Zeeman Instituut, Universiteit van Amsterdam, 
Valckenierstraat 65, 1018 XE Amsterdam, The Netherlands}

\author{Yvan Castin}
\email{yvan.castin@lkb.ens.fr}
\affiliation{Laboratoire Kastler Brossel, \'Ecole Normale
Sup\'erieure, 24 rue Lhomond, 75231 Paris Cedex 05, France}

\begin{abstract}
We show that a Bose-Einstein condensate with a vortex lattice in a
rotating anisotropic harmonic potential exhibits a very low frequency
scissors mode. The appearance of this mode is due to the SO(2)
symmetry-breaking introduced by the vortex lattice 
and, as a consequence of this,
its frequency tends to zero with the trap anisotropy $\epsilon$, with
a generic $\sqrt{\epsilon}$ dependence. We present analytical formulas
giving the mode frequency in the low $\epsilon$ limit and we show that
the mode frequency for some class of vortex lattices can tend to zero
as $\epsilon$ or faster.  We demonstrate that the standard classical
hydrodynamics approach fails to reproduce this low frequency mode,
because it does not contain the discrete structure of the vortex
lattice.
\end{abstract}



\date{\today}

\maketitle

Since their first observations in 1995 \cite{Cornell95,Ketterle95}, the trapped
gaseous atomic Bose-Einstein condensates have been the subject of intense
experimental \cite{revue_experiments} and theoretical \cite{revue_theory}
investigation.  A fruitful line of studies has been the measurement of the
eigenmode frequencies of the Bose-Einstein condensate, which can be performed
with a very high precision and therefore constitutes a stringent test of the
theory. Of particular interest in the context of superfluidity are the scissors
modes, first introduced in nuclear physics \cite{nuclear} and occurring in
quantum gases stored in a weakly anisotropic harmonic trap. These scissors modes
have been studied theoretically \cite{Odelin} and observed experimentally
\cite{Foot} in a non-rotating gas, both in a condensate (where they have a
large frequency even in the limit of a vanishing trap asymmetry) and in a
classical gas (where their frequency tends to zero in a vanishing trap
asymmetry limit).

Another fruitful line of studies has been the investigation of rotating
Bose-Einstein condensates: above a critical rotation frequency of the
anisotropic trap, it was observed experimentally that vortices enter the
condensate and settle into a regular lattice \cite{Dalibard,Ketterle,Cornell}.
In this paper, we study theoretically the scissors modes of a Bose-Einstein
condensate with several vortices present, a study motivated by the discovery of
a low frequency scissors mode in three-dimensional simulations in
\cite{crystal}.  In section \ref{sec:basic}, we define the model and the
proposed experiment, and we present results from a numerical integration of the
Gross-Pitaevskii equation:  condensates with several vortices present have a
scissors mode with a very low frequency, unlike non-rotating condensates.
We show in section \ref{sec:hydro} that the equations of classical hydrodynamics,
often used to describe condensates with a very large number of vortices
\cite{hydro_class}, fail to reproduce our numerical experiment. Then, in the
central part of our paper, section \ref{sec:Goldstone}, we interpret this mode
in terms of a rotational symmetry breaking due to the presence of the vortices,
which gives rise for $\epsilon=0$ to a Goldstone mode of zero energy, becoming
for finite $\epsilon$ a low frequency scissors mode and this leads to an
explanation for the failure of the classical hydrodynamics approximation.  We 
then use Bogoliubov theory in section \ref{sec:Bog} to obtain an upper
bound on the frequency on the scissors mode, which reveals that two situations
may occur: in what we call the non-degenerate case, the frequency of the
scissors mode tends to zero as the square root of the trap anisotropy, and we
show with perturbation theory that the upper bound gives the exact result in
this limit. In the opposite degenerate case, the frequency of the scissors mode
tends to zero at least linearly in the trap anisotropy, and this we illustrate
with a numerical calculation of the mode frequency for a three-vortex
configuration. We conclude in section \ref{sec:conclusion}.

\section{Model and a numerical simulation}
\label{sec:basic}

We consider a weakly interacting quasi two-dimensional (2D) Bose-Einstein condensate in an
anisotropic harmonic trap rotating at the angular frequency $\Omega$ \cite{quasi2D}. 
In the rotating frame,
the trapping potential is static and given by
\begin{equation}
U(x,y) = \frac{1}{2} m\omega^2 \left[(1-\epsilon) x^2 + (1+\epsilon) y^2\right]
\end{equation}
where $m$ is the mass of an atom, $\epsilon>0$ is the trap anisotropy and $\omega$
is the mean oscillation frequency of the atoms. From now on, we shall always remain
in the rotating frame. 

The condensate is initially in a stationary state of the Gross-Pitaevskii equation:
\begin{equation}
\left[-\frac{\hbar^2}{2m} \Delta + g|\psi|^2 + U - \Omega L_z -\mu\right] \psi= 0
\label{eq:gpe}
\end{equation}
where $g$ is the 2D coupling constant describing the atomic interactions \cite{Dum}, $\mu$
is the chemical potential, $\psi$ is the condensate field normalized to the number of particles
$N$ and $L_z=x p_y-y p_x$ is the angular momentum operator along $z$. In this paper, we shall
concentrate on the case that the rotation frequency $\Omega$ is large enough so that several
vortices are present in the field $\psi$, forming a regular array.

The standard procedure to excite the scissors mode is to rotate abruptly the trapping potential
by a small angle $\theta$ and to keep it stationary afterwards. This is theoretically
equivalent to abruptly rotating the field $\psi$ by the angle $-\theta$ while keeping
the trap unperturbed:
\begin{equation}
\psi(t=0^+) = e^{i \theta L_z/\hbar} \psi (t=0^-).
\label{eq:tourne}
\end{equation}
The subsequent evolution of $\psi$ is given by the time dependent Gross-Pitaevskii equation.

\begin{figure}[t]
\begin{tabular}{ccccc}
\includegraphics[width=3.0cm,height=3.0cm,clip=]{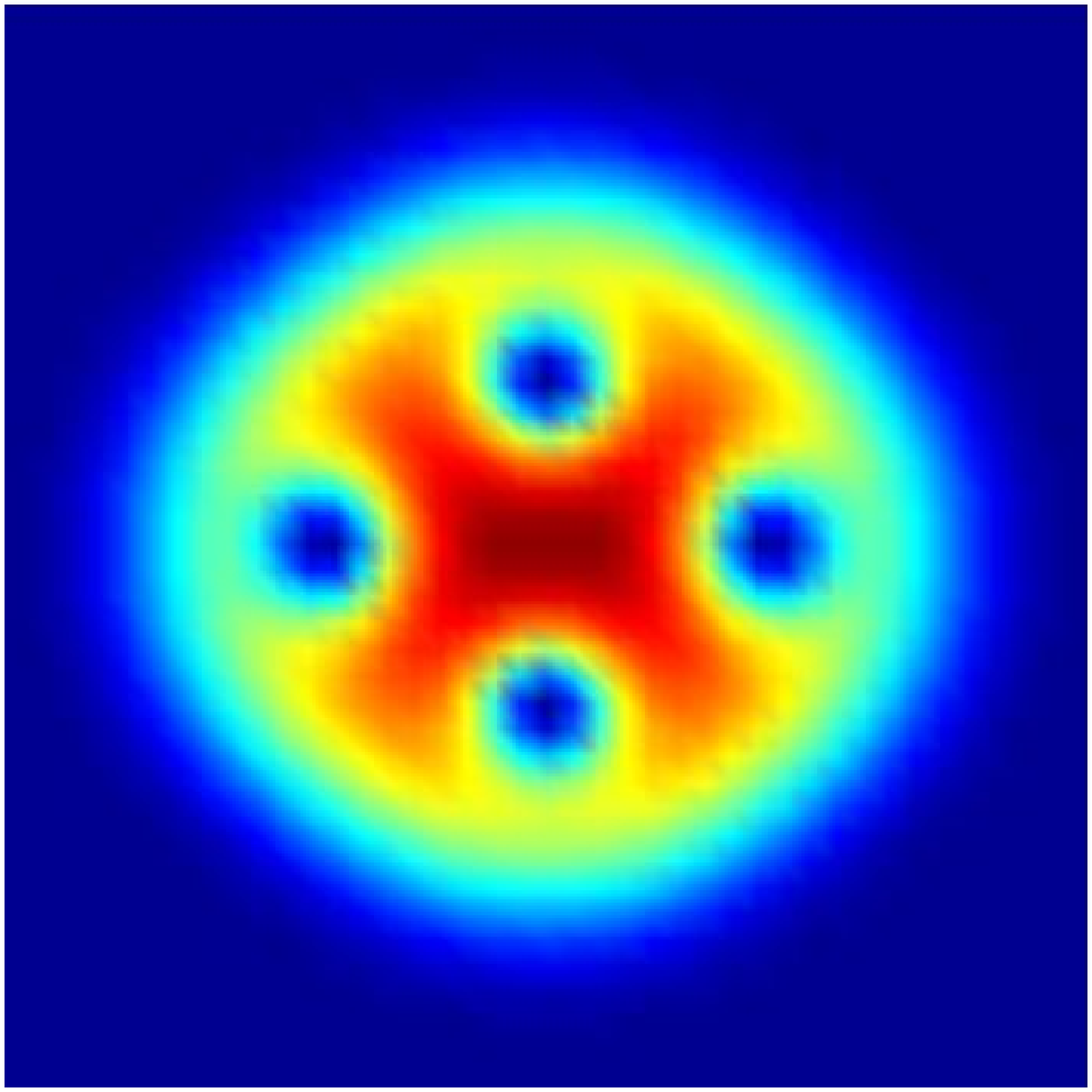}&
\includegraphics[width=3.0cm,height=3.0cm,clip=]{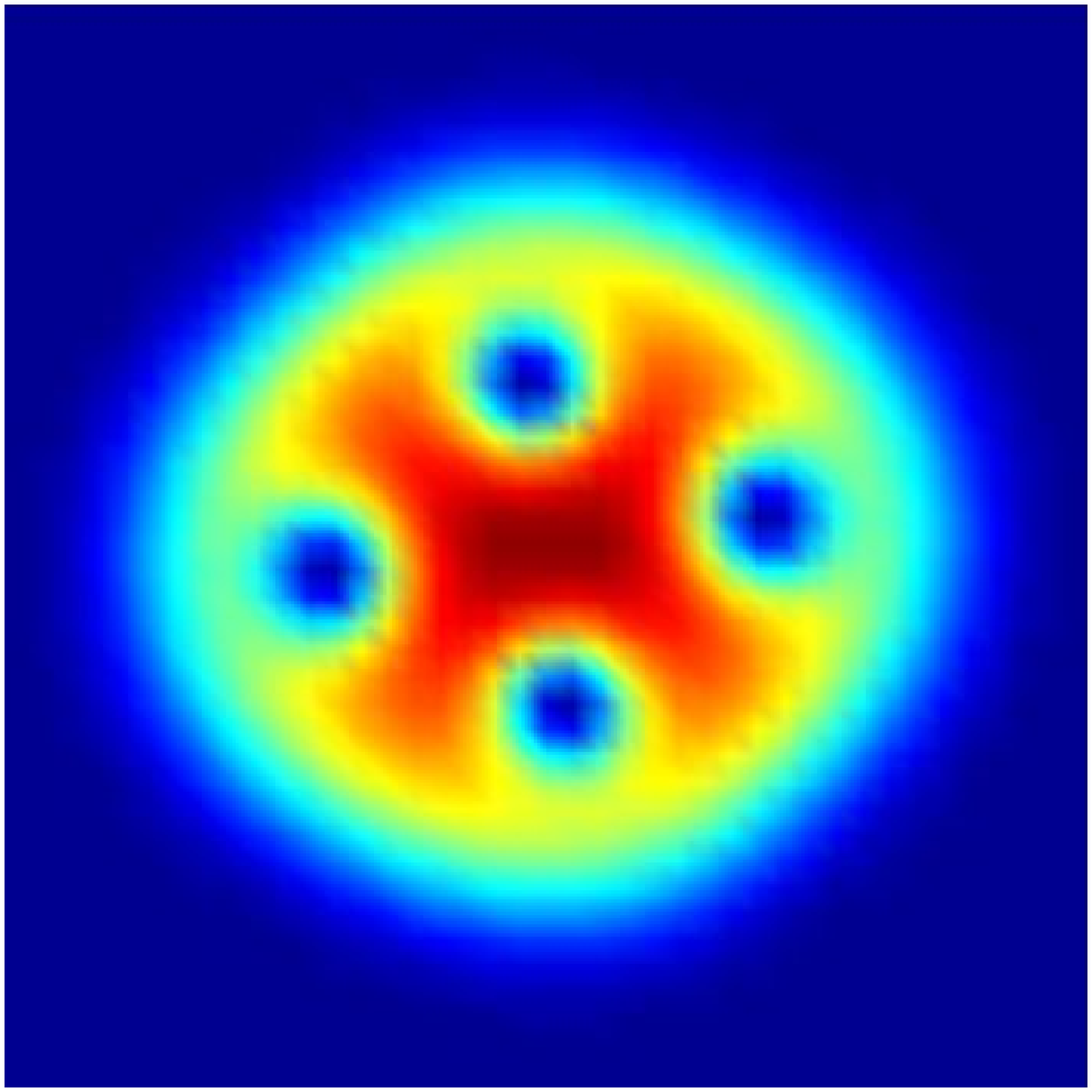}&
\includegraphics[width=3.0cm,height=3.0cm,clip=]{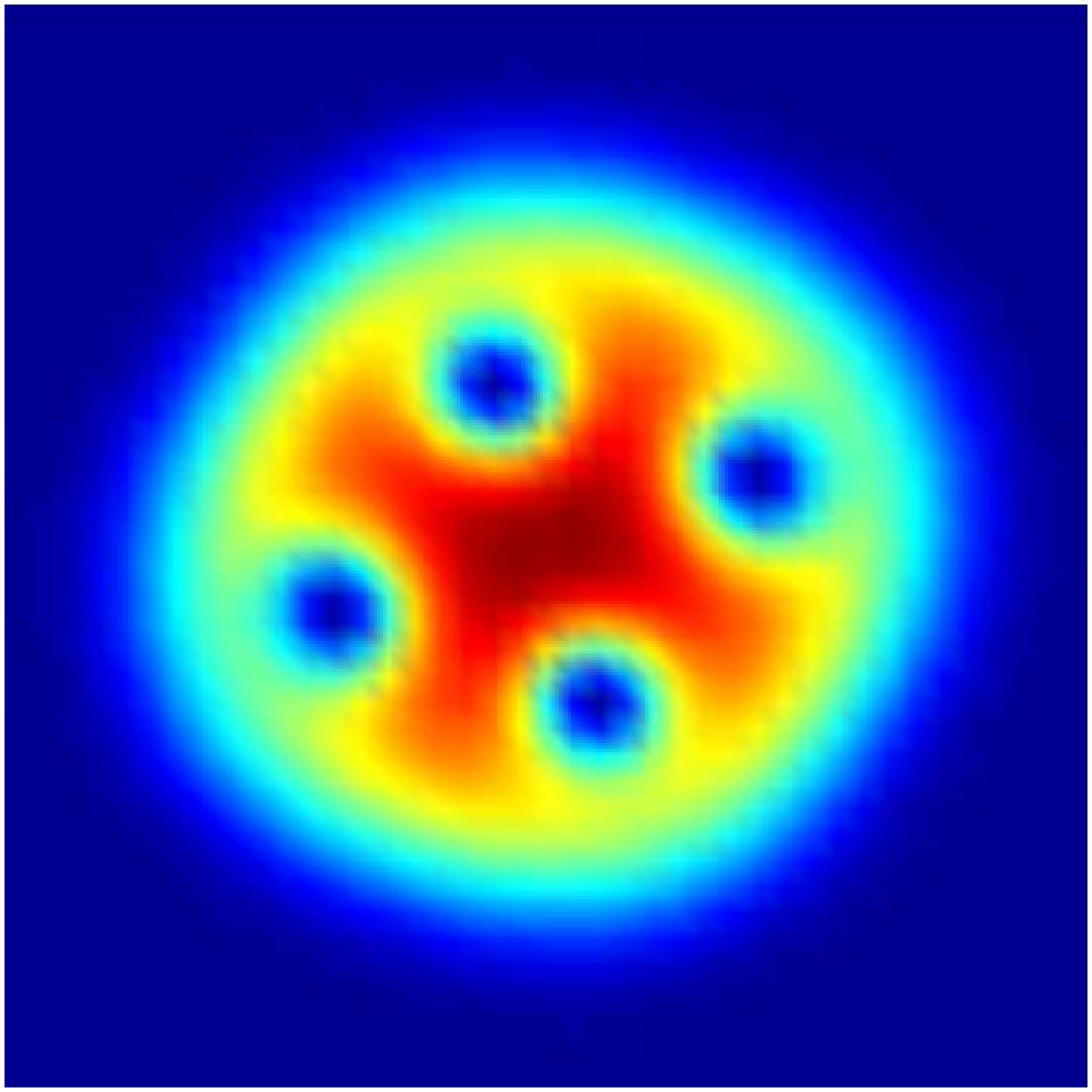}&
\includegraphics[width=3.0cm,height=3.0cm,clip=]{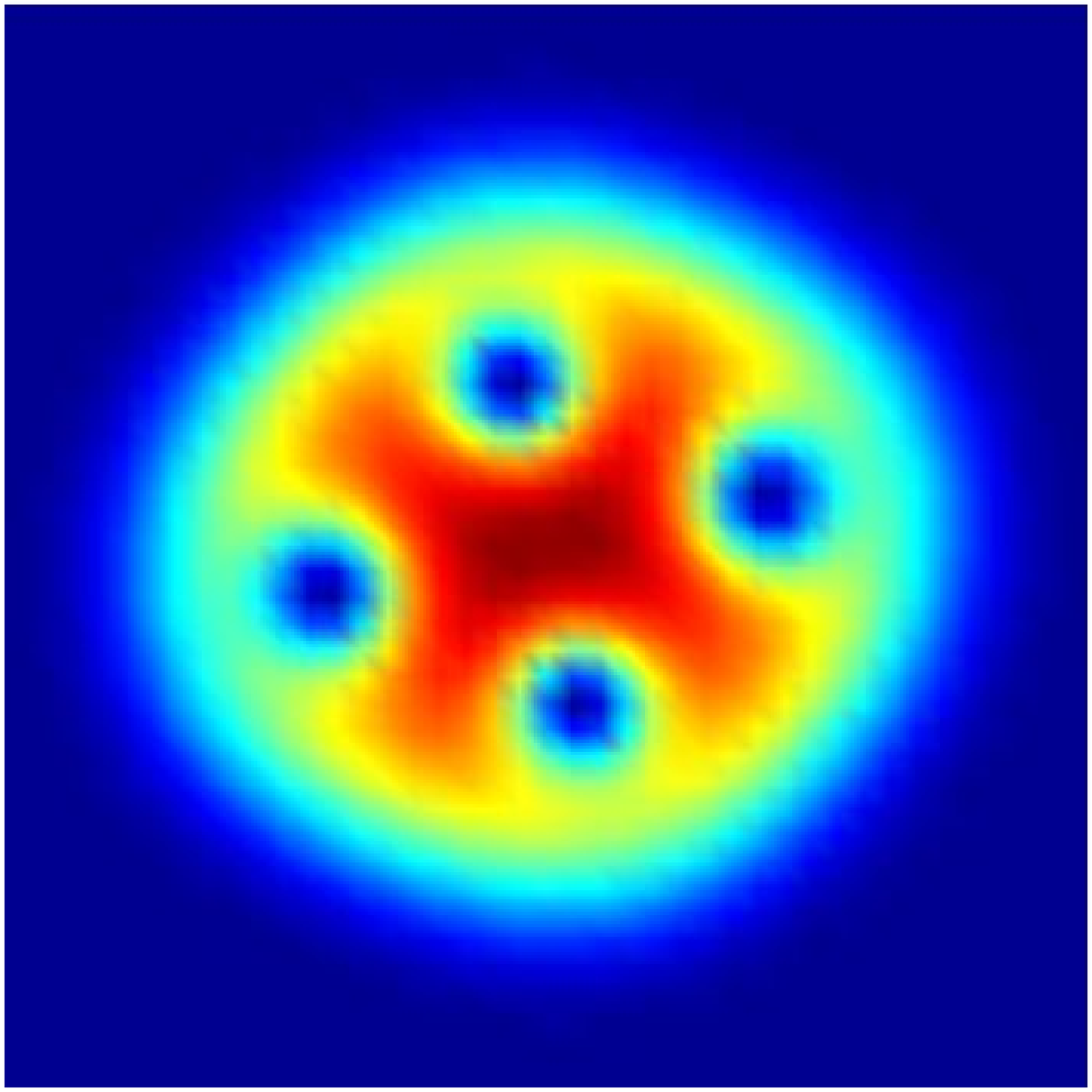}&
\includegraphics[width=3.0cm,height=3.0cm,clip=]{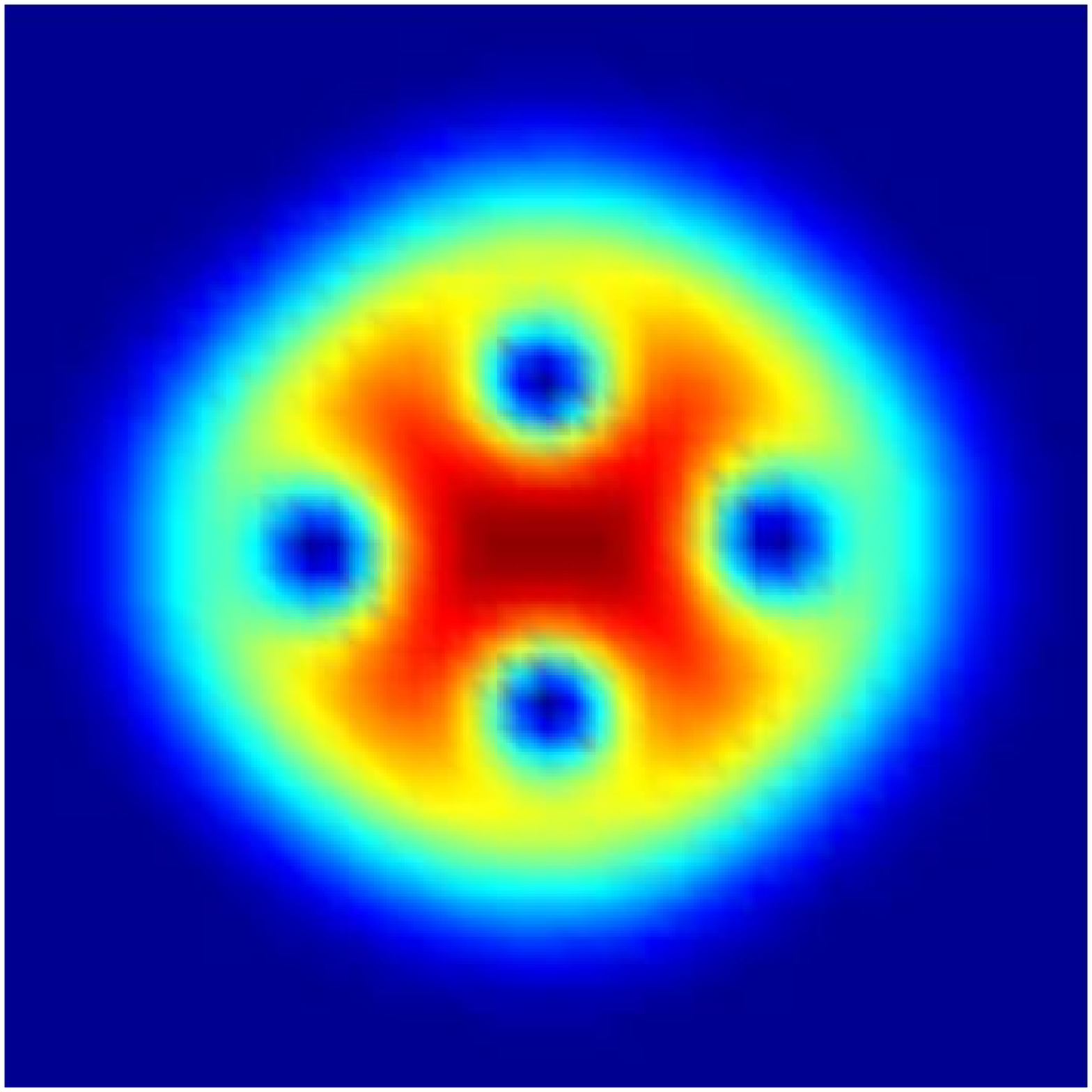} \\
\includegraphics[width=3.0cm,height=3.0cm,clip=]{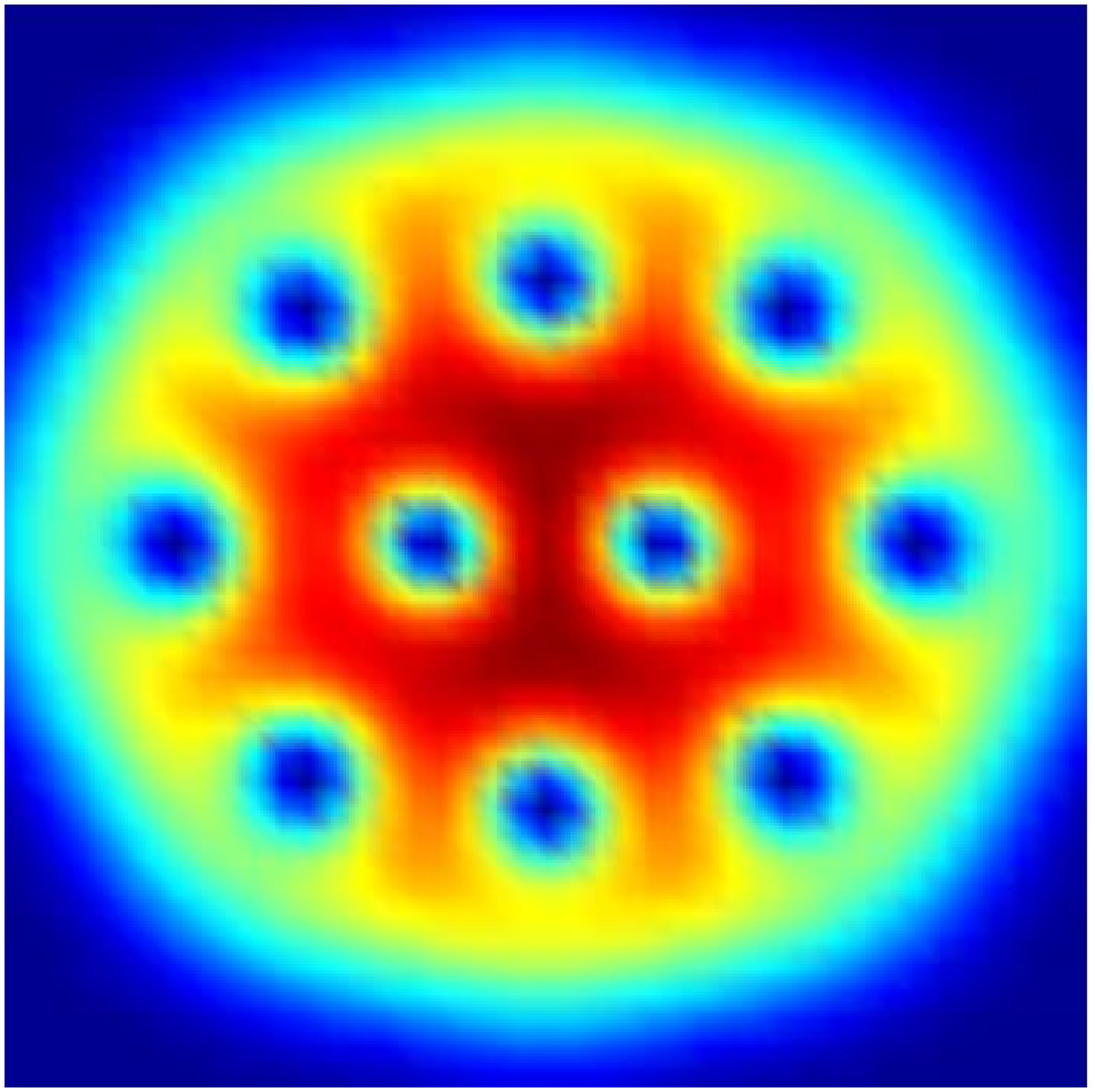}&
\includegraphics[width=3.0cm,height=3.0cm,clip=]{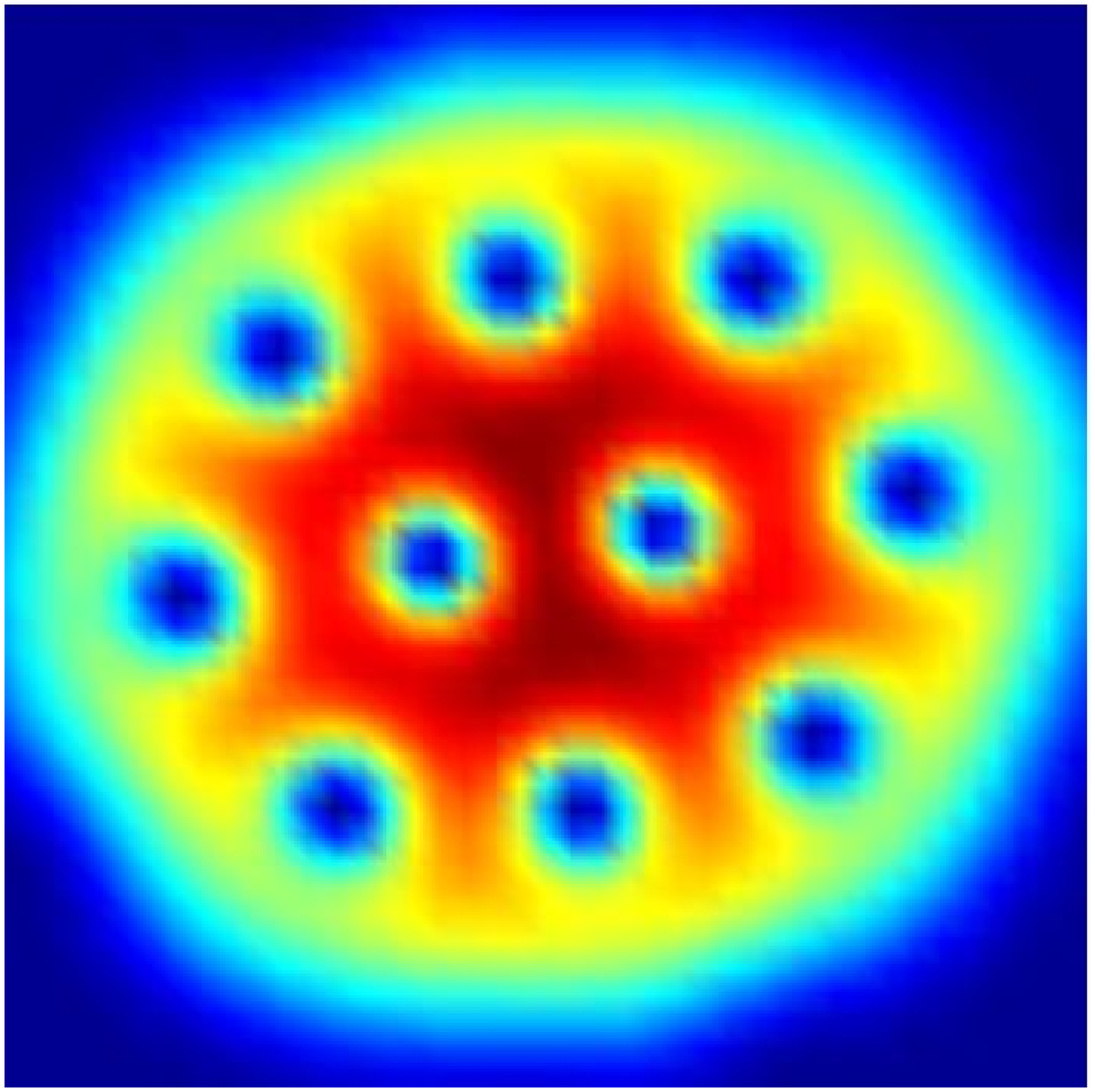}&
\includegraphics[width=3.0cm,height=3.0cm,clip=]{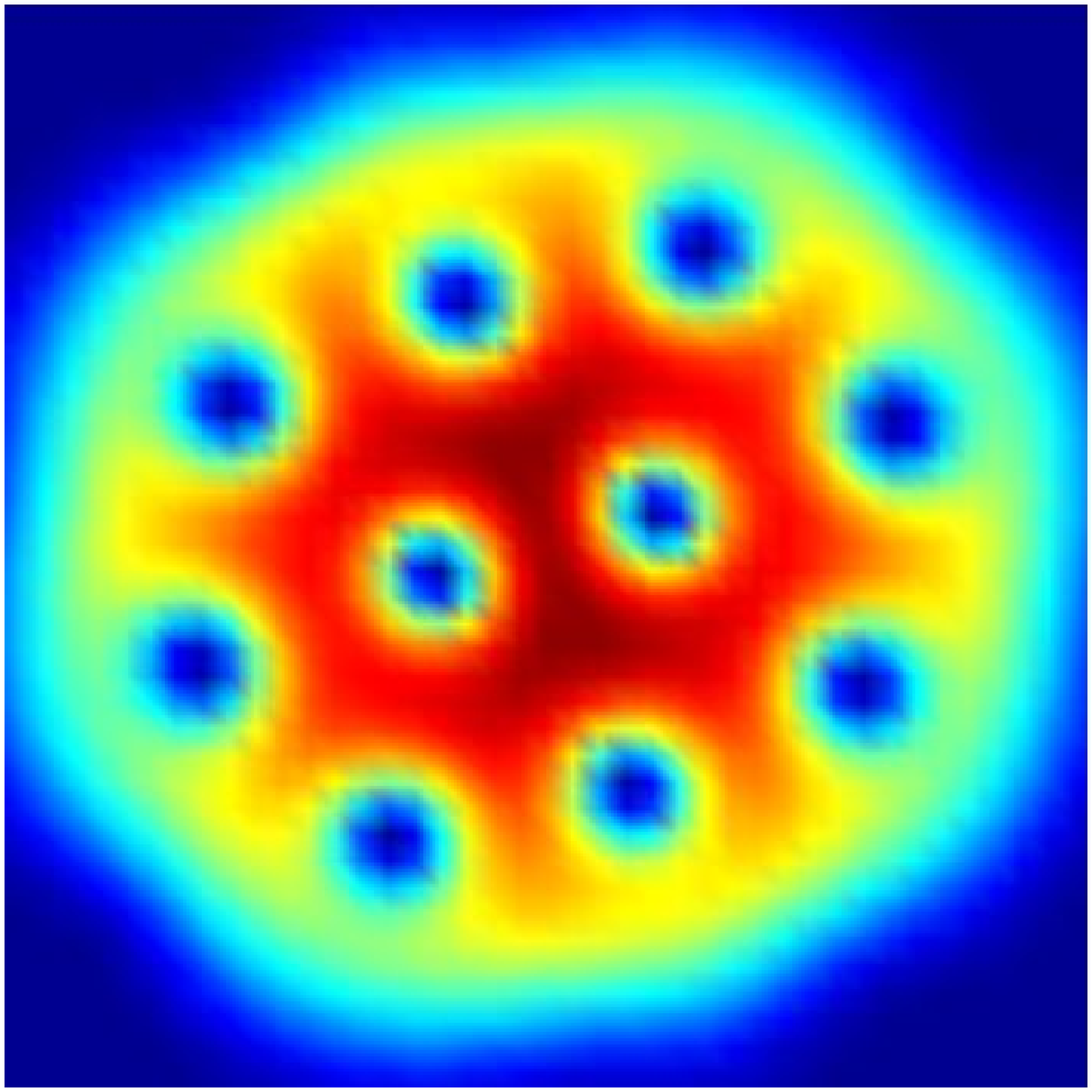}&
\includegraphics[width=3.0cm,height=3.0cm,clip=]{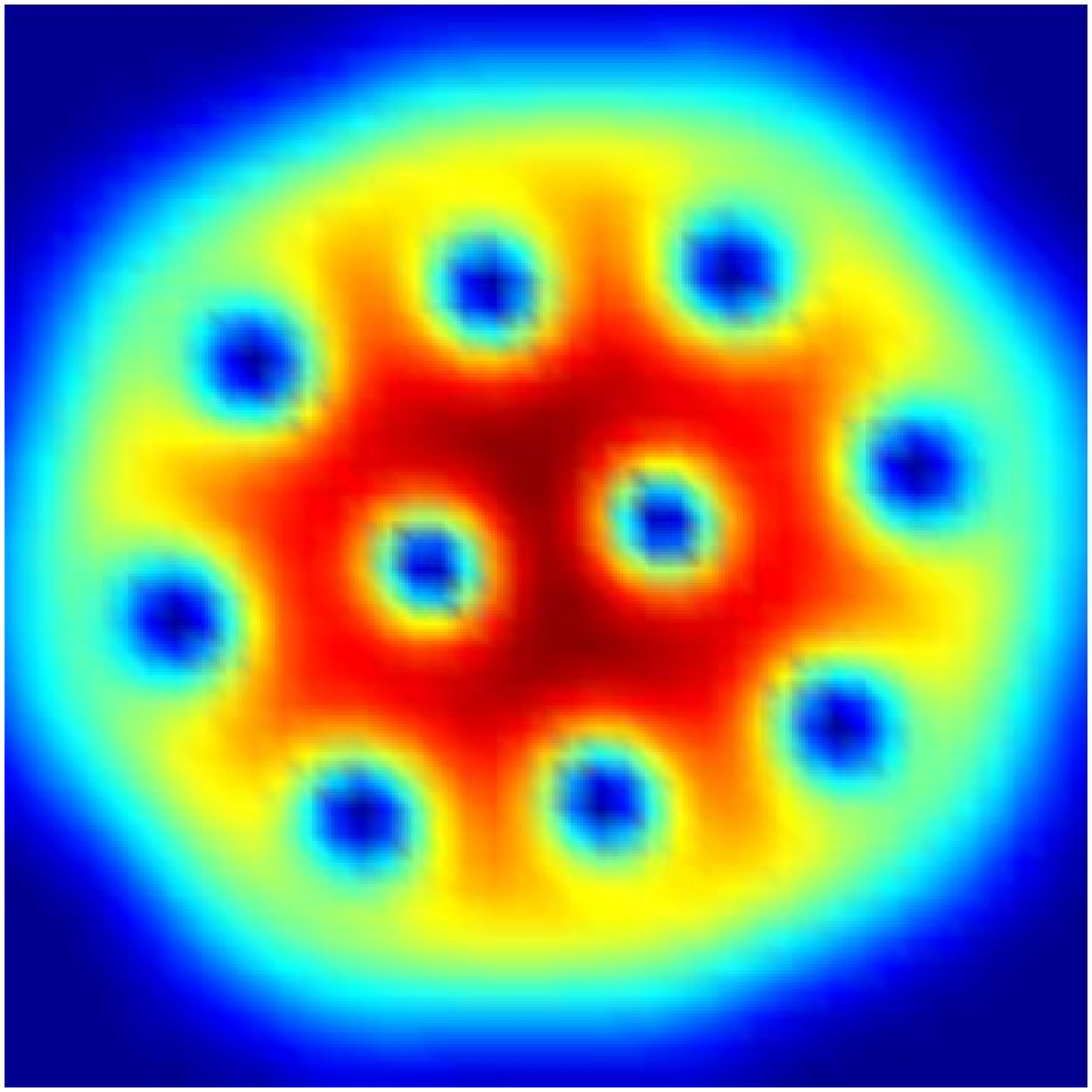}&
\includegraphics[width=3.0cm,height=3.0cm,clip=]{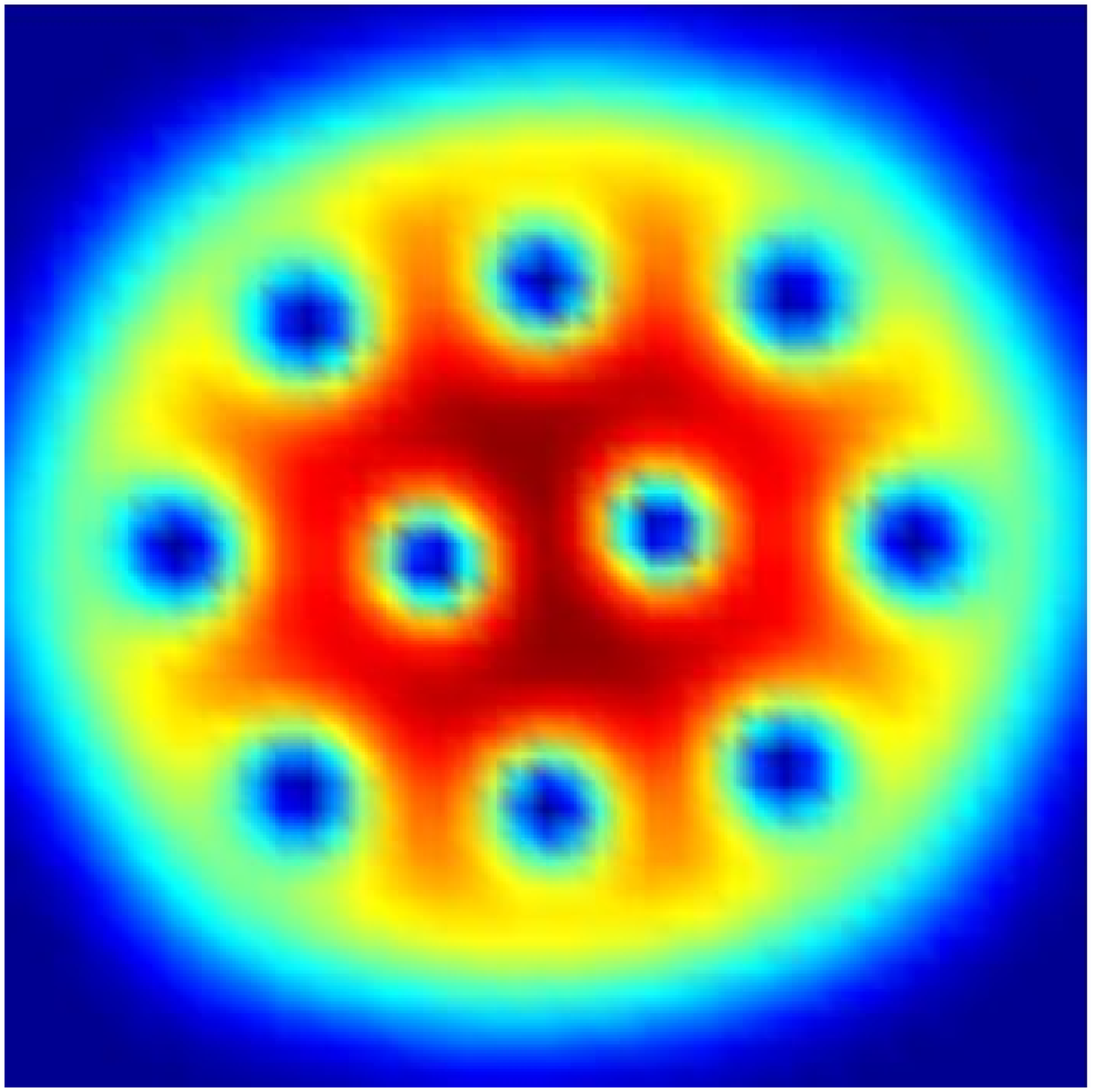}
\end{tabular}
\caption{For an initial stationary condensate with vortices, numerical simulation of the excitation
procedure of the scissors mode, by rotation of the trap by 10 degrees: density maps
at successive times. Top row: 4-vortex configuration, with $N=2500$ atoms and
a trap rotation frequency $\Omega=0.7\omega$, resulting in $\mu\simeq 4.6\hbar\omega$; the oscillation
period is $\sim 265  \omega^{-1}$ and the frames are separated by one quarter of a period.
Bottom row: 10-vortex configuration, with $N=6000$ atoms and
a trap rotation frequency $\Omega=0.8\omega$, resulting in $\mu\simeq 5.9\hbar\omega$; the oscillation
period is $\sim 245 \omega^{-1}$ and the frames are separated by one quarter of a period.
The trap anisotropy is $\epsilon=0.025$, the coupling constant is $g=0.0343\hbar^2/m$.
We have watched these oscillations without seeing significant damping over 7 periods.
\label{fig:simul}}
\end{figure}

We have solved this time dependent Gross-Pitaevskii equation numerically, with
the FFT splitting techniques detailed in \cite{Dum}, the initial state being
obtained by the conjugate gradient method \cite{Modugno}. The results for a
4-vortex and a 10-vortex configurations are shown in figure \ref{fig:simul} for
$\epsilon=0.025$, for an initial rotation of the trapping potential by an angle
of 10 degrees: we see a small amplitude slow oscillation of the vortex
lattice as a whole around the new axis of the trap, accompanied by weak internal
oscillations of the lattice and of the condensate. This suggests that, in this
low $\epsilon$ limit, the excitation procedure mainly excites the scissors mode,
and that this scissors mode has a very low frequency, much lower than the trap
frequency $\omega$.

\section{Classical hydrodynamics}
\label{sec:hydro}

In the approximation of a coarse-grained vorticity, a condensate with many 
vortices is described by a density profile
$\rho(\mathbf{r})$ and a velocity field $\mathbf{v}(\mathbf{r})$ solving the classical hydrodynamics equations:
\begin{eqnarray}
\label{eq:Euler}
\left(\partial_t +\mathbf{v}\cdot\mathbf{grad}\right) \mathbf{v} &=&
-\frac{1}{m}\mathbf{grad}\,(U+ \rho g) -2 \mathbf{\Omega}\times \mathbf{v} - 
\mathbf{\Omega}\times (\mathbf{\Omega}\times \mathbf{r} ) \\
\partial_t \rho + \mathrm{div}\, \rho \mathbf{v} &=& 0.
\end{eqnarray}
The first equation is Euler's equation in the rotating frame, including the Coriolis
and centrifugal forces. The second one is the continuity equation. 

The velocity field $\mathbf{v}$ is the velocity field of the fluid in the rotating frame.
In a stationary state, we set $\mathbf{v}=\mathbf{0}$ which amounts to assuming 
the solid body velocity
field $\mathbf{\Omega}\times \mathbf{r} $ in the lab frame. The corresponding stationary density profile
is given by a quadratic ansatz:
\begin{equation}
\rho_0(x,y) = \frac{m}{g}(\mu -a_1 x^2 -a_2 y^2).
\end{equation}
From Eq.(\ref{eq:Euler}) one then finds
\begin{equation}
a_{1,2} = \frac{1}{2} \left[\omega^2(1\mp\epsilon) -\Omega^2\right].
\end{equation}

What happens after the rotation of the density profile by a small angle? To answer this question 
analytically, we linearize the hydrodynamics equations around the steady state, setting
$\rho=\rho_0+\delta\rho$ and $\mathbf{v}= \delta\mathbf{v}$:
\begin{eqnarray}
\label{eq:Euler_lin}
\partial_t \delta\mathbf{v} &=&
-\frac{1}{m}\mathbf{grad}\,\delta\rho g -2 \mathbf{\Omega}\times \delta\mathbf{v} \\
\partial_t \delta\rho + \mathrm{div}\, \rho_0 \delta\mathbf{v} &=& 0.
\end{eqnarray}
At time $t=0^+$, $\delta\mathbf{v} =0$ and, to first order in the rotation angle $\theta$,
$\delta\rho = 2 \theta m x y (a_1-a_2)/g$. The subsequent evolution is given by the time
dependent polynomial ansatz:
\begin{eqnarray}
\delta\rho &=& \frac{m}{g}\left[c(t)-\mathbf{r}\cdot \delta A(t) \mathbf{r}\right] \\
\delta\mathbf{v} &=& \delta B(t) \mathbf{r}
\end{eqnarray}
where $\delta A(t)$ is a time-dependent 2$\times$2 symmetric matrix and
$\delta B(t)$ is a time-dependent 2$\times$2 general matrix. These 
matrices
evolve according to
\begin{eqnarray}
\label{eq:dA}
\delta\dot{A} &=& - A \mathrm{Tr} \delta B(t) - A \delta B(t) - \delta B^T(t) A \\
\delta\dot{B} &=& 2 \, \delta A(t) + 2 \Omega \begin{pmatrix} 
0 & 1 \\ -1 & 0\end{pmatrix} \delta B(t)
\label{eq:dB}
\end{eqnarray}
where we have set $A=\begin{pmatrix} a_1 & 0 \\ 0 & a_2 \end{pmatrix}$. The constant
term evolves as $\dot{c}+\mu {\rm Tr}\delta B=0$.

First, one may look for eigenfrequencies of the system solved by $\delta A$ and
$\delta B$. For $\epsilon=0$, this may be done analytically \cite{Cozzini}: one finds 
one mode of zero frequency, and 6 modes of non-zero frequencies:
\begin{equation}
\nu = \pm 2\omega, \pm \left[(2\omega^2-\Omega^2)^{1/2}+\Omega\right], 
\pm \left[(2\omega^2-\Omega^2)^{1/2}-\Omega\right].
\label{eq:hydro_nu}
\end{equation}
At this stage, the presence of a zero energy mode for vanishing anisotropy looks
promising to explain the low frequency of the scissors mode. 
At weak but non-zero $\epsilon$, however, a numerical diagonalization of the 
resulting 7$\times$7 matrix shows that the zero frequency is unchanged,
whereas the others change very little. Analytically, one then easily finds the zero-frequency
mode for arbitrary $\epsilon$:
\begin{equation}
\delta A = -\Omega A \ \ \ \ \ \ \delta B = \begin{pmatrix} 0 & -a_2 \\ a_1 & 0 \end{pmatrix}
\end{equation}

Why is there such a zero-frequency mode in the classical hydrodynamics? 
We have found that this is because there a continuous branch of stationary 
solutions
of the classical hydrodynamics equations parametrized by two real numbers $\alpha$
and $\beta$:
\begin{eqnarray}
\mathbf{v} &=& \begin{pmatrix} \alpha y \\ \beta x \end{pmatrix} \\
g\rho/m &=&  \mu -a_+ x^2 -a_- y^2
\end{eqnarray}
where
\begin{eqnarray}
a_+ &=& a_1-\Omega\beta+\frac{1}{2}\alpha\beta \\
a_- &=& a_2+\Omega\alpha+\frac{1}{2}\alpha\beta.
\end{eqnarray}
These real numbers are not independent since they have to satisfy
\begin{equation}
\alpha \beta (\alpha+\beta) + 2\alpha a_1 +2 \beta a_2 = 0.
\end{equation}
This is a second degree equation for $\beta$ at a given $\alpha$ so it can be
solved explicitly, giving rise to two branches. One of them 
contains the $\mathbf{v}=\mathbf{0}$  
stationary
solution as a particular case, with $\alpha=\beta=0$; it terminates in a point where $a_+=a_-=0$.
Each stationary solution on this branch has a zero-frequency mode.

\begin{figure}[htb]
\includegraphics[height=5.0cm,clip=]{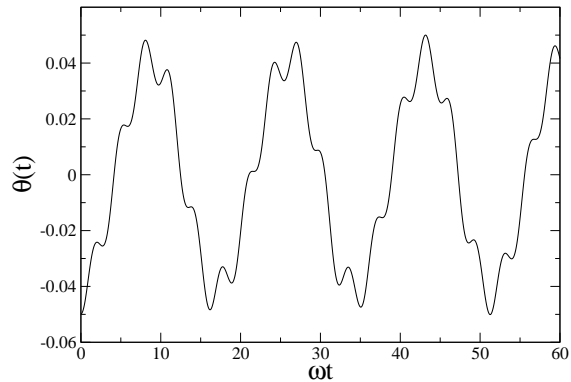}
\caption{In the linearized classical hydrodynamics approximation, 
angle of the long axis of the condensate with the $x$ axis of the trap, as a function
of time, after the stationary density profile was turned abruptly by a small angle. Here
$\epsilon=0.025$ and $\Omega=0.8\omega$. No low frequency oscillation appears: 
the main oscillation frequency is $\sim 0.37 \omega$, in agreement with the
lowest frequency of the $\epsilon=0$ limit given by Eq.(\ref{eq:hydro_nu}). }
\label{fig:classical}
\end{figure}

So what happens after the scissors mode excitation, in the classical hydrodynamics approximation?
We have numerically integrated the linearized equations Eqs.(\ref{eq:dA},\ref{eq:dB}) with the initial
conditions specified above these equations (see figure \ref{fig:classical}). We find a scissors mode oscillation, however at a large
frequency close to the $\epsilon=0$ prediction $\nu=\left[(2\omega^2-\Omega^2)^{1/2}-\Omega\right]$, in
disagreement with the simulations of the previous section \cite{note_pour_carlos}.

\section{Rotational symmetry breaking: from Goldstone mode to scissors mode}
\label{sec:Goldstone}

Having failed to find the scissors mode of section \ref{sec:basic} using the 
classical hydrodynamics approximation we now discard it and turn to a quantum 
treatment of the problem.
Consider a stationary solution of the Gross-Pitaevskii equation with a vortex lattice 
present, in the case of an isotropic trap, $\epsilon=0$. 
This solution clearly breaks the rotational symmetry SO(2) of the Hamiltonian.
Since this is a continuous symmetry group, Goldstone's theorem guarantees us the existence
of a degree of freedom behaving as a massive boson \cite{Blaizot}. In 
particular, this implies the
existence of a zero energy eigenmode of the condensate, corresponding to the rotation of
the wave function in real space. The ``mass" of the Goldstone boson can be shown to be
\begin{equation}
\label{eq:masse}
M= \partial_\Omega \int \psi^* L_z \psi.
\end{equation}
The variable $\theta$ conjugate to the Goldstone momentum $P$ has the physical meaning of
being a rotation angle of the lattice with respect to a reference direction.
Note that a quantum spreading of this angular variable takes place when time proceeds,
as a consequence of the Goldstone Hamiltonian $P^2/2M$.

Now, for a non-zero value of the trap anisotropy, there is no SO(2) symmetry any longer in the
Hamiltonian, so that the Goldstone mode is turned into a regular mode of the condensate, behaving as a
harmonic oscillator with a finite frequency. This means that the variable $\theta$ now experiences
a potential. An intuitive estimate of this potential is obtained by taking the stationary
solution, rotating it by an angle $\theta$ and calculating its Gross-Pitaevskii energy in the presence
of a trap anisotropy:
\begin{equation}
{\cal H} = \frac{P^2}{2M} + V(\theta) 
\end{equation}
where we keep only the $\theta$ dependent part of the energy:
\begin{equation}
V(\theta) = \frac{1}{2} m \omega^2 \epsilon \left[\langle x^2-y^2\rangle (1-\cos 2\theta) +2 \langle xy\rangle \sin 2\theta\right]
\end{equation}
where $\langle \ldots\rangle= \int \psi^* \ldots \psi$.
We note in passing that the fact that $\psi$ is a local minimum of energy imposes that $V(\theta)\geq 0$ for all
rotation angle $\theta$ so that 
\begin{eqnarray}
\langle xy\rangle &=& 0 \\ 
\langle x^2-y^2\rangle &\geq& 0 .
\end{eqnarray}
Quadratising $V(\theta)$ around $\theta=0$ leads to the prediction that the angle 
$\langle \theta \rangle$ oscillates when initially its value is different from 
zero. The resulting motion of the condensate is an oscillation of the lattice 
as a whole, i.\ e.\ a scissors mode with an angular frequency:
\begin{equation}
\nu_{\rm scissors}^2 \simeq \epsilon \frac{2m \omega^2\langle x^2-y^2\rangle}{\partial_\Omega \langle L_z\rangle}.
\label{eq:nu1}
\end{equation}
This results in a $\sqrt{\epsilon}$ dependence of the mode frequency in the case when
$\langle x^2-y^2\rangle$ does not vanish in the low $\epsilon$ limit.

It is now clear why classical hydrodynamics does not exhibit this low energy 
mode: since it does not break SO(2) symmetry, the corresponding Goldstone mode 
is absent and so the low energy scissors mode does not appear.
We note that, in cases where SO(2) symmetry breaking would occur in the hydrodynamics equations,
the scissors mode frequency would be predicted to tend to zero, even if the motion of the
particles is not quantized; this was indeed shown by an explicit
solution of the superfluid hydrodynamic equation for a rotating
vortex-free condensate exhibiting rotational
symmetry breaking and the $\sqrt{\epsilon}$ dependence of the scissors mode
frequency was seen experimentally \cite{Jean_et_Sandro}.

\section{Analytical results from Bogoliubov theory}
\label{sec:Bog}

\subsection{Technicalities of Bogoliubov theory}

A systematic way of calculating the eigenfrequencies of the condensate is simply to linearize
the time dependent Gross-Pitaevskii equation around a stationary solution, taking as unknown
the vector $(\delta \psi,\delta\psi ^*)$ where $\delta\psi$ is the deviation of the field from
its stationary value. One then
faces the following operator,
\begin{equation}
{\cal L}_{\rm GP} = \begin{pmatrix}
-\frac{\hbar^2}{2m}\Delta + U + 2g |\psi|^2 -\Omega L_z -\mu & g \psi^2  \\
-  g \psi^{*2} & -\left[-\frac{\hbar^2}{2m}\Delta + U + 2g |\psi|^2 -\Omega L_z -\mu\right]^*
\end{pmatrix}
\end{equation}
where the complex conjugate of an operator $X$ is defined as $\langle \mathbf{r} |X^*|\mathbf{r}'\rangle
=\langle \mathbf{r} |X|\mathbf{r}'\rangle^*$.
The scissors mode we are looking for is an eigenstate of this operator, and the corresponding
eigenvalue over $\hbar$ gives the frequency of the mode.

At this stage, a technical problem appears, coming from the fact that the operator
${\cal L}_{\rm GP}$ is not diagonalizable \cite{CastinDum}. This can be circumvented by using
the number-conserving Bogoliubov theory, in which the eigenmodes of the condensate are the eigenvectors
of the following operator:
\begin{equation}
{\cal L} =  
\begin{pmatrix}Q & 0 \\ 0 & Q^* \end{pmatrix}
{\cal L}_{\rm GP}
\begin{pmatrix}Q & 0 \\ 0 & Q^* \end{pmatrix}
\end{equation}
where $Q$ projects orthogonally to the condensate wave function 
\begin{equation}
\phi=\frac{\psi}{N^{1/2}}, 
\end{equation}
and $Q^*$ therefore projects orthogonally to $\phi^*$.
We shall assume here that the operator ${\cal L}$
is diagonalizable for non-zero values of the trap anisotropy
$\epsilon$. We shall see in the next subsection that it is in general not diagonalizable
for $\epsilon=0$ when several vortices are present.

One then introduces the eigenmodes of $\cal L$, written as
$(u_k,v_k)$, such that $\int |u_k|^2-|v_k|^2=1$; to each of these
eigenmodes of eigenvalue $\epsilon_k$  corresponds an eigenmode of $\cal L$ of eigenvalue $-\epsilon_k^*$, given
by $(v_k^*,u_k^*)$ \cite{Houches}. We shall assume here that the condensate wave function is a 
dynamically
stable solution of the stationary Gross-Pitaevskii equation, so that all the $\epsilon_k$ are real.
We shall also assume that the condensate wave function is a local minimum of the 
Gross-Pitaevskii
energy functional (condition of thermodynamical metastability) so that all the $\epsilon_k$
are positive.

\subsection{An upper bound for the lowest energy Bogoliubov mode}
\label{subsec:lowest}

We can then use the following result. For a given deviation $\delta\psi^*$ of the condensate
field from its stationary value, we form the vectors
\begin{equation}
\vec{e}_q \equiv {\cal L}^q \begin{pmatrix} Q \delta\psi \\ Q^*\delta\psi^* \end{pmatrix}
\end{equation}
where $q$ is an integer.  As shown in the appendix \ref{appen:bound} 
we then have the following inequality:
\begin{equation}
\mathrm{min}_k\, \epsilon_k \leq \left(\frac{\langle \vec{e}_0,\vec{e}_1\rangle}
{\langle \vec{e}_0,\vec{e}_{-1}\rangle}\right)^{1/2}
\label{eq:inegal}
\end{equation}
where the scalar product $\langle , \rangle$ is of signature $(1,-1)$:
\begin{equation}
\langle \begin{pmatrix}u \\v\end{pmatrix}, \begin{pmatrix}u'\\v'\end{pmatrix}\rangle 
= \int u^* u' - v^* v'.
\label{eq:scalar}
\end{equation}
We apply this inequality, taking for $\delta \psi$ the deviation originating from the rotation of
$\psi$ by an infinitesimal angle: expanding Eq.(\ref{eq:tourne}) to first order in the rotation
angle, we put 
\begin{equation}
\vec{e}_0 =\begin{pmatrix} i Q L_z \psi \\-i Q^* L_z^* \psi^* \end{pmatrix}. \label{eq:e0} 
\end{equation}
After lengthy calculations detailed in the
appendix \ref{appen:bound},
we obtain the upper bound
\begin{equation}
\mathrm{min}_k\, \epsilon_k^2 \leq \frac{\hbar^2 \langle \phi| (\partial_\theta^2 U)|\phi\rangle}
{\partial_\Omega(\langle \phi| L_z |\phi\rangle)}
\label{eq:upper}
\end{equation}
where $\theta$ is the angle of polar coordinates in the $x-y$ plane.

In the limit of a small $\epsilon$, we assume that the scissors mode is the lowest energy Bogoliubov mode,
as motivated in section \ref{sec:Goldstone}. Eq.(\ref{eq:upper}) then gives an
upper bound to the scissors mode frequency.  An explicit calculation of $\partial_\theta^2 U$
shows that the upper bound Eq.(\ref{eq:upper}) coincides with the intuitive estimate
Eq.(\ref{eq:nu1}):
\begin{equation}
\partial_\theta^2 U = 2 m \omega^2 \epsilon \langle \phi| x^2-y^2 |\phi \rangle.
\end{equation}

This upper bound suggests two possibilities in the low $\epsilon$ limit. In the first one,
$\langle \phi| x^2-y^2 |\phi \rangle$ tends to a non-zero value, which implies that the scissors mode
frequency vanishes at most as $\epsilon^{1/2}$. In the second possibility, 
$\langle \phi| x^2-y^2 |\phi \rangle$ tends to zero; under the reasonable assumption of
a $\langle \phi| x^2-y^2 |\phi \rangle$ vanishing linearly with $\epsilon$, this implies a
scissors mode frequency vanishing at most as $\epsilon$. This second case we term `degenerate'.

We point out that the degenerate case contains all the cases where the stationary wave function 
$\psi$
at vanishing trap anisotropy has a discrete rotational symmetry with an angle $\gamma$ different from
$\pi$. To show this, we introduce the coordinates rotated by an angle $\gamma$,
\begin{eqnarray}
x' &=& x \cos\gamma -y \sin \gamma \\
y' &=& x \sin\gamma +y \cos \gamma.
\end{eqnarray}
The symmetry of $\psi$ implies that $\langle \phi| x^2-y^2 |\phi \rangle = \langle \phi| x'^{2}-y'^{2} |\phi \rangle$
and $\langle \phi| xy |\phi \rangle = \langle \phi| x'y' |\phi \rangle$. By expansion of these identities, and
assuming $1-\cos 2\gamma \neq 0$, one gets $\langle \phi| x^2-y^2 |\phi \rangle=0$.

An important question is to estimate the relative importance of the modes
other than the scissors mode excited by the sudden rotation of
the condensate. As shown in the appendix \ref{appen:bound}, under
the assumption that the scissors mode is the only mode of vanishing frequency
when $\epsilon\rightarrow 0$, and anticipating some results of the
next subsection, the weight of the initial excitation
$\vec{e}_0$ on the non-scissors modes behaves as
\begin{equation}
||\vec{e}_0^{\, \rm non-scissors}||^2 = O(\epsilon^2)
\end{equation}
both in the degenerate and the non-degenerate cases.

\subsection{Perturbation theory in $\epsilon$}

We now treat the anisotropic part of the trapping potential as a perturbation, $\delta U = m \omega^2 \epsilon
(y^2-x^2)/2$. The Bogoliubov operator $\cal L$, considered as a function of $\epsilon$,
can be written as
\begin{equation}
{\cal L}_\epsilon = {\cal L}_0 + \delta {\cal L}
\end{equation}
where $\delta {\cal L}$ is a perturbation. Note that the explicit expression of $\delta {\cal L}$ cannot be
given easily, as it involves not only $\delta U$ but also the effect of the first order change of the condensate
wave function entering the mean field terms of $\cal L$; but we shall not need such an explicit 
expression.

Apart from eigenmodes of non-zero eigenfrequency, the operator ${\cal L}_0$ has a normal zero energy mode
characterized by the vector $\vec{e}_n$ and an anomalous mode characterized
by the vector $\vec{e}_a$, in accordance with Goldstone's theorem. They are given by
\begin{eqnarray}
\label{eq:en}
\vec{e}_n &=& \begin{pmatrix} Q_0 L_z \psi_0 \\ -Q_0^* L_z^* \psi_0^*  \end{pmatrix} \\
\vec{e}_a &=& \begin{pmatrix} Q_0 \partial_\Omega \psi_0 \\ Q_0^* \partial_\Omega \psi_0^*  \end{pmatrix} 
\label{eq:ea}
\end{eqnarray}
where $\psi_0$ is the stationary condensate field for $\epsilon=0$ and
$Q_0$ is the projector on the $\epsilon=0$ condensate wavefunction
$\phi_0$. As shown in the appendix
\ref{appen:bound}, one has indeed ${\cal L}_0 \vec{e}_n =0$ and
${\cal L}_0 \vec{e}_a = \vec{e}_n$. Within the subspace generated by 
$\vec{e}_n$ and $\vec{e}_a$, the operator ${\cal L}_0$ has therefore the Jordan canonical form:
\begin{equation}
{\cal L}_0|_{\rm subspace} = \begin{pmatrix}  0 & 1 \\ 0 & 0 \end{pmatrix}.
\end{equation}
Note that the normal and the anomalous vectors are, up to a global factor, adjoint vectors for the modified
scalar product Eq.(\ref{eq:scalar}), in the sense that
$\langle \vec{e}_n,\vec{e}_a\rangle = M_0$ where $M_0$ is given in Eq.(\ref{eq:masse}) for $\epsilon=0$.

As usual in first order perturbation theory, one takes the restriction of the perturbation $\delta {\cal L}$
to the subspace and one then diagonalizes it: 
\begin{equation}
{\cal L}|_{\rm subspace} = \begin{pmatrix}  \delta{\cal L}_{nn} & 1+\delta{\cal L}_{na}  \\ \delta{\cal L}_{an} & 
\delta{\cal L}_{aa} \end{pmatrix}.
\label{eq:2par2}
\end{equation}
The notation $\delta{\cal L}_{an}$ means than one takes the component of $\delta{\cal L} \vec{e}_n$ onto
the vector $\vec{e}_a$. Since the adjoint vector of $\vec{e}_a$ is $\vec{e}_n/M_0$ one has
\begin{equation}
\delta{\cal L}_{an} = M_0^{-1}\langle \vec{e}_n, \delta{\cal L} \vec{e}_n\rangle.
\end{equation}
A similar notation is used for $\delta{\cal L}_{nn}$ and
$\delta{\cal L}_{aa}$.
Forming the characteristic polynomial of the 2$\times$2 matrix Eq.(\ref{eq:2par2}) and taking all the $\delta{\cal L}$'s to be
$O(\epsilon)$, one realizes that to leading
order in $\epsilon$, the eigenvalues are $\pm \delta{\cal L}_{an}^{1/2}$, that is they scale as $\epsilon^{1/2}$
if $\delta{\cal L}_{an}\neq 0$.

Finally, we use the exact identity $\langle \vec{e_0}, {\cal L} \vec{e_0}\rangle= N \langle \phi| (\partial_\theta^2 U)
|\phi \rangle$, proved in the appendix \ref{appen:bound}; noting that $\vec{e}_n$ and $\vec{e}_0$ differ by
terms of first order in $\epsilon$, expanding ${\cal L}$ in ${\cal L}_0 + \delta {\cal L}$ and
using the fact that $\langle \vec{e}, {\cal L}_0 \vec{e}_n\rangle = \langle \vec{e}_n, {\cal L}_0 \vec{e}\rangle=0$
whatever the vector $\vec{e}$, we get for the scissors mode angular frequency
\begin{equation}
\nu_{\rm scissors}^2 = \delta{\cal L}_{an}+ O(\epsilon^2) = 
\frac{\langle \phi_0| (\partial_\theta^2 U) |\phi_0 \rangle}{\partial_\Omega 
\langle\phi_0|L_z|\phi_0\rangle}+ O(\epsilon^2). 
\label{eq:nusquared}
\end{equation}
When $\langle x^2-y^2\rangle_0 \neq 0$,
this coincides with the upper bound Eq.(\ref{eq:upper}) to leading order in $\epsilon$: 
the upper bound is then saturated for low $\epsilon$.

\subsection{Analytic expressions in the Thomas-Fermi limit for the non-degenerate case}

In the Thomas-Fermi limit ($\mu \gg \hbar\omega$) there exist asymptotic
functionals giving the energy of the vortex lattice as a function of the vortex positions 
in the isotropic \cite{Dum} and anisotropic \cite{Aftalion1} cases. Here we shall use 
these functionals to evaluate the frequency of the scissors mode in 
Eq.(\ref{eq:nusquared}).

From the general expression (Eq.(2.12) in \cite{Aftalion1}) we take
the simplifying assumption that the distances of the vortex cores to the
trap center are much smaller that the Thomas-Fermi radius of the condensate,
to obtain
\begin{eqnarray}
\label{eq:aftalion}
\Delta E &=& \Delta E_0 +\frac{\eta^2m \omega^2}{2} \left( \sum_i (1-\epsilon) x_i^2+(1+\epsilon) 
y_i^2\right) \left( -2 \pi | \log \eta | + 2\sqrt{2\pi} \frac{\Omega}{\eta 
\omega}\right) \nonumber \\
&&-\sqrt{2\pi} \hbar \omega \eta \sum_{i\neq j} \log (d_{ij}/l) + O(\epsilon^2)
\end{eqnarray}
where $\Delta E$ is the energy difference per atom
between the configurations with and without 
vortices, $\eta=\hbar/\sqrt{2Nmg}=\hbar\omega/(\mu \sqrt{2\pi}) \ll 1$ (with $\mu$ being
the Thomas-Fermi chemical potential of the non-rotating gas), 
$(x_i,y_i)$ are the coordinates of the i${}^{\rm th}$ vortex 
core, $d_{ij}$ is the distance between the vortex cores $i$ and $j$ and $l$ 
is a length scale on the order of the Thomas-Fermi radius
and independent of $\epsilon$ \cite{tech2}. 
$\Delta E_0$ is a quantity independent of $\epsilon$ and 
of the vortex positions:
therefore we shall not need its explicit expression.
It is assumed that $\Omega\ll\omega$, and that
\begin{equation}
\Omega > \Omega_m = \left(\frac{\pi}{2}\right)^{1/2}\eta\omega | \log \eta |  
\end{equation}
i.e.\ the rotation frequency is large enough to ensure that each vortex experiences
an effective trapping potential close to the trap center.
An important property of the simplified energy functional is that
the positions of the vortices minimizing Eq.(\ref{eq:aftalion})
are universal quantities not depending on $\Omega$ 
\cite{Aftalion1} when they are rescaled
by the length $\lambda$ \cite{o_vs_om}: 
\begin{equation}
\lambda^2 = \frac{\hbar}{2m(\Omega-\Omega_m)}.
\end{equation}

Now we calculate $\langle x^2-y^2 \rangle$ using the Hellman-Feynman theorem:
\begin{equation}
\frac{d}{d\epsilon}E=-\frac{1}{2}m\omega^2 \int(x^2-y^2) |\phi|^2
\end{equation}
where $E$ is the energy per atom and $\phi$ is the condensate wave function. We have 
also that
\begin{equation}
\frac{d}{d\epsilon}E(\epsilon=0)=\frac{d}{d\epsilon}\Delta E(\epsilon=0)
\end{equation}
since the vortex-free configuration is rotationally symmetric for
$\epsilon=0$.
We thus obtain \cite{really_technical}
\begin{equation}
\lim_{\epsilon \rightarrow 0} \langle \phi|(x^2-y^2)|\phi \rangle=2\sqrt{2\pi}
\eta  \frac{\Omega-\Omega_m}{\omega}
\sum_i x^2_i-y^2_i.
\label{eq:x2my2_asympt}
\end{equation}

\begin{figure}[htb]
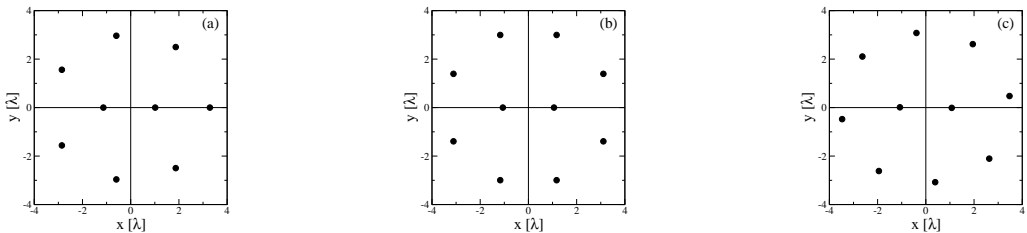

\includegraphics[width=3.0cm,height=3.0cm,clip=]{uni_9.eps}
\hfil
\includegraphics[width=3.0cm,height=3.0cm,clip=]{uni_10.eps}
\hfil
\includegraphics[width=3.0cm,height=3.0cm,clip=]{uni_10_inter.eps}
\caption{Using the simplified energy functional 
Eq.(\ref{eq:aftalion}) we find the minimum energy configurations 
of nine vortices (a) and ten vortices (b) for $\epsilon=0$. These configurations are non-degenerate: 
$\sum_i (x_i^2-y_i^2)/\sum_i (x_i^2+y_i^2) \simeq 0.03$
for both of them. For $0.0173<\epsilon<0.0182$ a cross-over takes
place between the 10-vortex configuration of (a) and the one of
figure \ref{fig:simul}: such a cross-over configuration is shown in (c),
for $\epsilon=0.018$.}
\label{fig:toto}
\end{figure}

Using this formula we study vortex lattices with up to ten vortices. 
By considering first the case $\epsilon=0$, we find that they are
all degenerate except the case of two, nine and ten vortices, see figure \ref{fig:toto}a for $n_v=9$ and figure \ref{fig:toto}b for $n_v=10$.
Interestingly the $\epsilon=0$ 
10-vortex configuration differs from the one of the full numerical
simulation of figure \ref{fig:simul}. By minimizing the simplified energy functional for a
non-zero $\epsilon$, we find that the 10-vortex configuration of figure \ref{fig:simul}
ceases indeed 
to be a local minimum of energy when $\epsilon$ is smaller than 0.0182: 
this
prevents us from applying Eq.(\ref{eq:nusquared}) of perturbation theory to the
calculation of the scissors mode frequency of the numerical simulations.
The 10-vortex configuration of figure \ref{fig:toto}b ceases to be a local
minimum of energy when $\epsilon$ is larger than 0.0173. 
The 10-vortex configuration
minimizing the energy for $0.0173 < \epsilon < 0.0182$ is a distorted
configuration which breaks both the $x$ and $y$ reflection
symmetries of the
energy functional, but not the parity invariance, 
see figure \ref{fig:toto}c.

The last point is to calculate the derivative of the mean angular momentum with respect
to $\Omega$. In the Thomas-Fermi limit, for $\epsilon=0$, and to first order in the squared
distance of the vortex cores from the trap center, the angular momentum
is given by \cite{Dum,Aftalion1}:
\begin{equation}
\lim_{\epsilon \rightarrow 0} \langle L_z\rangle \simeq \hbar n_v  - \sqrt{2\pi}
m\omega\eta \sum_i x_i^2 + y_i^2,
\end{equation}
where $n_v$ is the number of vortices. 
Using the rescaling by $\lambda$ to isolate the $\Omega$ dependence
we get
\begin{equation}
\frac{d}{d\Omega} \langle L_z\rangle(\epsilon=0) = - \sqrt{2\pi}{m\omega\eta}
\frac{d\lambda^2/d\Omega}{\lambda^2} \sum_{i} x_i^2 + y_i^2
\end{equation}
where $d\lambda^2/d\Omega = -\lambda^2/(\Omega-\Omega_m)<0$.
In this way, $\nu_{\rm scissors}$ for low $\epsilon$ can be calculated 
from Eq.(\ref{eq:nusquared}) in the non-degenerate case analytically
in terms of the equilibrium positions of the vortices:
\begin{equation}
\nu_{\rm scissors}^2 = 4\epsilon (\Omega-\Omega_m)^2
\frac{\sum_i x_i^2-y_i^2}{\sum_i x_i^2+y_i^2}.
\label{eq:full_analy}
\end{equation}

\subsection{Numerical results in the degenerate case}

\begin{figure}[htb]
\includegraphics[height=5.0cm,clip=]{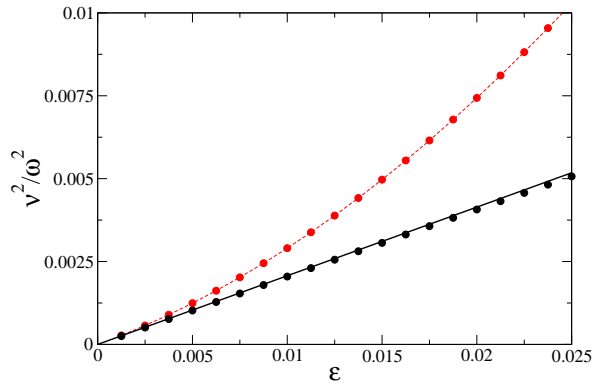}
\caption{Frequency squared of the scissors mode as a function of 
the trap anisotropy $\epsilon$ for a condensate with two vortices. $N=2500$, $\Omega=0.6 
\omega$ and $g=0.0343\hbar^2/m$ resulting in a chemical potential $\mu \simeq 4.9 \hbar \omega$. 
The black circles are determined from the iteration 
of the operator ${\cal L}^{-1}$. The red disks correspond to the numerical
evaluation of the upper bound Eq.(\ref{eq:upper}).
The solid line corresponds to perturbation theory:
it is the numerical evaluation of the leading term in $\epsilon$
of Eq.(\ref{eq:nusquared}), which is linear in $\epsilon$ with a non-zero
slope since the two-vortex configuration is non-degenerate.
The red dashed line is a guide to the eyes.
\label{fig:cis3}}
\end{figure}

\begin{figure}[htb]
\includegraphics[height=5.0cm,clip=]{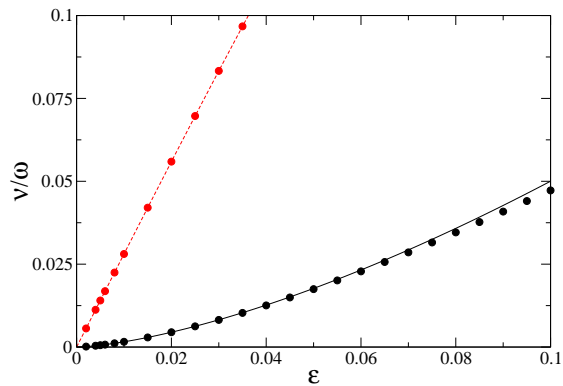}
\caption{Frequency of the scissors mode as a function of
the trap anisotropy $\epsilon$ for a condensate with three vortices. $N=2500$, $\Omega=0.7
\omega$ and $g=0.0343\hbar^2/m$ resulting in a chemical potential $\mu \simeq 4.6 \hbar \omega$. 
The black circles are determined from the iteration
of the operator ${\cal L}^{-1}$. The red disks correspond to the numerical
evaluation of the upper bound Eq.(\ref{eq:upper}).
The result of perturbation theory is not shown, since it vanishes for
a degenerate configuration.
The black line is given by the equation $\nu=1.582 \epsilon^{3/2}$ and results
from a fit of the black disks by a power law $\epsilon^{3/2}$ over the interval
$\epsilon\leq 0.01$.
The red dashed line is a guide to the eyes.
\label{fig:cis14a16}}
\end{figure}

In the degenerate case we must go to higher order in perturbation 
theory to get the leading order prediction for the frequency of the scissors 
mode. A simpler alternative is to calculate numerically the frequency of this 
mode by iterating the operator ${\cal L}^{-1}$ \cite{Brachet} starting with 
the initial guess $\vec{e}_0$ defined in Eq.(\ref{eq:e0}) \cite{tech}.
The corresponding numerical results are presented in figure \ref{fig:cis3} in the
non-degenerate case of two vortices, and in figure \ref{fig:cis14a16} in the
degenerate case of three vortices. In the non-degenerate case, $\nu_{\rm scissors}^2$
is found to scale linearly with $\epsilon$ for low $\epsilon$, as expected, and the
corresponding slope is is very good agreement with the prediction
Eq.(\ref{eq:nusquared}); the agreement with the asymptotic formula
Eq.(\ref{eq:full_analy}) is poor, which could be expected
since the parameters are not deeply enough in the Thomas-Fermi
regime \cite{o_vs_om}. In the degenerate case with three vortices, we find
numerically that $\nu_{\rm scissors}$ scales as $\epsilon^{3/2}$ for low
$\epsilon$, which is indeed compatible with Eq.(\ref{eq:upper})
which leads to a scissor frequency upper bound scaling as $\epsilon$.
The fact that a strictly higher exponent ($3/2>1$) is obtained shows that some
cancellation happens in the next order of perturbation theory, may be
due to the threefold symmetry of the vortex configuration.

\section{Conclusion}
\label{sec:conclusion}

In this paper we have studied the problem of a non-classical scissors mode of a
condensate containing a vortex lattice. In 2D simulations of the
Gross-Pitaevskii equation we showed that, when such a condensate experiences a
sudden rotation by a small angle of the anisotropic harmonic trapping
potential, the orientation of its vortex lattice will undergo very low
frequency oscillations of the scissors mode type. Motivated by these numerical
results we searched for this mode using the well-known classical hydrodynamics
approximation where the condensate density is a smooth function showing no sign
of the presence of vortices and where their effect on the velocity field is
taken into account through a coarse-grained vorticity. We showed that this
approximation does not contain a low energy scissors mode. We then were able to
explain this discrepancy using the Gross-Pitaevskii equation that treats
quantum mechanically the motion of the particles. In this case, the density
profile of the vortex lattice breaks the rotational symmetry which naturally
gives rise to a Goldstone mode in the limit of an isotropic trap and, for a
finite anisotropy it becomes a low energy scissors mode.
The existence of a scissors mode at a low value of the mode frequency 
is therefore a direct consequence of the discrete nature
of the vortices in a condensate, which is itself a consequence of the quantization
of the motion of the particles.

We obtained quantitative predictions for the mode frequency using two separate methods. First we
calculated an upper bound on the frequency of the mode using an inequality
involving the Bogoliubov energy spectrum. Second, using perturbation theory in
$\epsilon$, the anisotropy of the trapping harmonic potential, we showed that
this inequality becomes an equality to leading order in $\epsilon$ when the
expectation value $\langle x^2-y^2 \rangle_0$ taken in the unperturbed state
does not vanish; in this case the frequency of the scissors mode tends to zero
as $\epsilon^{1/2}$, and we gave an analytic prediction for the coefficient in front
of $\epsilon^{1/2}$ in the Thomas-Fermi limit. However, in the cases where the expectation value does
vanish (which we termed ``degenerate''), the frequency will be at most linear 
in $\epsilon$. We have illustrated this using a three-vortex lattice where the 
frequency vanishes as $\epsilon^{3/2}$.
Also, in the general case, we have shown that the relative weight
of the non-scissors modes excited by a sudden infinitesimal 
rotation of the trap
tends to zero as $O(\epsilon^2)$ so that the excitation procedure
produces a pure scissors mode in the low trap anisotropy limit.

Finally we point out that the existence of the non-classical scissors mode does not rely on the
fact that we are dealing with a Bose-Einstein condensate {\it per
se}. In particular, we expect, based on the very general arguments of
section \ref{sec:Goldstone}, that this mode will equally be present
in a Fermi superfluid containing a vortex lattice and that its frequency
will tend to zero as $\sqrt{\epsilon}$ in the non-degenerate case just
like in its bosonic counterpart.

Laboratoire Kastler Brossel is a research unit of Ecole normale
sup\'erieure and Universit\'e Paris 6, associated to CNRS.
This work is part of the research program on quantum gases of the
Stichting voor Fundamenteel Onderzoek der Materie (FOM), which is
financially supported by the Nederlandse Organisatie voor Wetenschappelijk
Onderzoek (NWO).

\appendix
\section{Derivation of the upper bound on the Bogoliubov energies}
\label{appen:bound}

In this appendix, we prove several results used to control analytically
the frequency and the excitation weight of the scissors mode.

We first demonstrate the inequality Eq.(\ref{eq:inegal}). The key assumption
is that the the condensate wavefunction is a local
minimum of energy and does not break any continuous symmetry (which implies
here a non-zero trap anisotropy $\epsilon \neq 0$).
Then the Bogoliubov operator $\cal L$ of the number-conserving Bogoliubov
theory is generically expected to be diagonalizable, with all eigenvalues being real and non-zero.
We then expand $\vec{e}_{0}$ on the eigenmodes of $\cal L$:
\begin{equation}
\vec{e}_{0} = \sum_{k\in {\cal F}_+} b_k 
\begin{pmatrix}u_k \\ v_k \end{pmatrix}+
b_k^* \begin{pmatrix}v_k^* \\ u_k^* \end{pmatrix}
\end{equation}
where ${\cal F}_+$ is the set of modes normalized as $\int |u_k|^2-|v_k|^2 =1$,
and which have here strictly positive energies $\epsilon_k>0$ since
the condensate wavefunction is a local minimum of energy
\cite{Houches}. The $b_k$'s are complex numbers; the amplitudes on the modes
of the ${\cal F}_-$ family (of energies $-\epsilon_k$) are simply $b_k^*$ since $\vec{e}_{0}$ is of
the form $(f,f^*)$. Using the fact that for the modified scalar product
Eq.(\ref{eq:scalar}),
different eigenmodes are orthogonal, and each eigenmode has a `norm squared' equal
to $+1$ for ${\cal F}_+$ and $-1$ for ${\cal F}_-$, we get:
\begin{equation}
\frac{\langle \vec{e}_0,\vec{e}_1\rangle}
{\langle \vec{e}_0,\vec{e}_{-1}\rangle} = 
\frac{\sum_{k\in {\cal F}_+} 2 \epsilon_k |b_k|^2}{\sum_{k\in {\cal F}_+} 2 \epsilon_k^{-1} |b_k|^2}.
\end{equation}
This is simply the expectation value of $\epsilon_k^2$  with the positive weights
$\epsilon_k^{-1} |b_k|^2$. Hence Eq.(\ref{eq:inegal}).

Next, starting from $\vec{e}_0$ given by Eq.(\ref{eq:e0}), we have to calculate 
$\vec{e}_{-1}$ and $\vec{e}_1$ to obtain Eq.(\ref{eq:upper}).

To calculate $\vec{e}_{-1}$, we take the derivative of the Gross-Pitaevskii
equation Eq.(\ref{eq:gpe}) with respect to the rotation frequency
$\Omega$, which leads to
\begin{equation}
{\cal L}_{\rm GP} \begin{pmatrix} \partial_\Omega \psi \\
\partial_\Omega \psi^*\end{pmatrix} =
(\partial_\Omega \mu) \begin{pmatrix} \psi \\
-\psi^*\end{pmatrix}
+
\begin{pmatrix} L_z \psi \\
-L_z^* \psi^*\end{pmatrix}.
\end{equation}
We have to get $\cal L$ instead of ${\cal L}_{\rm GP}$: since 
$\psi$ is normalized to the number of particles $N$, which does not
depend on $\Omega$, we have
\begin{equation}
\int \psi^* \partial_\Omega \psi = i \gamma
\end{equation}
where $\gamma$ is real, so that 
$\partial_\Omega \psi = i \gamma \psi /N + Q (\partial_\Omega \psi)$
where $Q$ projects orthogonally to $\psi$. Using the fact
that $(\psi, -\psi^*)$ is in the kernel of ${\cal L}_{\rm GP}$,
we obtain:
\begin{equation}
\vec{e}_{-1} = \begin{pmatrix} i Q (\partial_\Omega \psi) \\
i Q^* (\partial_\Omega \psi^*)\end{pmatrix}
\label{eq:e-1}
\end{equation}
which allows to get
\begin{equation}
\langle \vec{e}_0,\vec{e}_{-1}\rangle=
\partial_\Omega \int \psi^* L_z \psi.
\label{eq:rule-1}
\end{equation}
In passing, we note that the fact that all $\epsilon_k$ are positive
implies that $\langle\vec{e}_0,\vec{e}_{-1}\rangle >0$
so that the mean angular momentum is an increasing function of $\Omega$,
a standard thermodynamic stability constraint for a system in contact
with a reservoir of angular momentum \cite{Rokhsar}.

We now proceed with the calculation of $\vec{e}_1\equiv {\cal L} \vec{e}_0$.
Since $L_z$ is hermitian, 
\begin{equation}
L_z \psi =  Q ( L_z \psi) - \alpha \psi
\end{equation}
where $\alpha$ is real. Since $(\psi, -\psi^*)$ 
is in the kernel of ${\cal L}_{\rm GP}$, we conclude that
\begin{equation}
{\cal L}_{\rm GP} \vec{e}_0 =
i \left[ {\cal L}_{\rm GP}, \begin{pmatrix} L_z & 0 \\
0 &  L_z^* \end{pmatrix} \right]
\begin{pmatrix} \psi \\ -\psi^* \end{pmatrix}
\end{equation}
where $[,]$ stands for the commutator.
One calculates the commutator and then computes its action
on $(\psi,-\psi^*)$: various simplifications occur so that
\begin{equation}
{\cal L}_{\rm GP} \vec{e}_0 = \begin{pmatrix} -\hbar (\partial_\theta U)
\psi \\ \hbar (\partial_\theta U) \psi^*
\label{eq:lgp_e0}
\end{pmatrix}
\end{equation}
where $\theta$ is the angle of polar coordinates in the $x-y$ plane
and where we used $L_z = -i \hbar \partial_\theta$.
$\vec{e}_1$ results from this expression by projection.
We finally obtain
\begin{equation}
\langle \vec{e}_0, \vec{e}_1\rangle =
\hbar^2 \int \psi^* (\partial_\theta^2 U) \psi.
\label{eq:rule1}
\end{equation}

In what concerns the scissors mode frequency,
the last point is to justify the identities ${\cal L}_0 \vec{e}_n=0$
and ${\cal L}_0\vec{e}_a = \vec{e}_n$, where ${\cal L}_0$ is the
$\epsilon=0$ limit of ${\cal L}$, $\vec{e}_n$
is defined in Eq.(\ref{eq:en}) and $\vec{e}_a$ is defined in Eq.(\ref{eq:ea}).
First, one notes that, within a global factor $i$,
$\vec{e}_n$ is the limit of $\vec{e}_0$ for $\epsilon\rightarrow 0$.
Then taking the limit $\epsilon\rightarrow 0$ in Eq.(\ref{eq:lgp_e0}),
one immediately gets ${\cal L}_0 \vec{e}_n=0$ since $\partial_\theta U$
vanishes when the trapping potential $U$ is rotationally symmetric.
Second, one notes from Eq.(\ref{eq:e-1}) that, within a global factor $i$,
$\vec{e}_a$ is the zero $\epsilon$ limit of $\vec{e}_{-1}$.
Then taking the zero $\epsilon$ limit of the identity
${\cal L} \vec{e}_{-1} = \vec{e}_0$ leads to ${\cal L}_0\vec{e}_a = \vec{e}_n$.

Finally we control the weight of the modes other than the scissors
mode that are excited by the sudden infinitesimal rotation
of the condensate: 
\begin{equation}
||\vec{e}_0^{\, \rm non-scissors}||^2 \equiv
||\sum_{k\in{\cal F}_+\setminus \{k_s\}} b_k
\begin{pmatrix}u_k \\ v_k \end{pmatrix}+
b_k^* \begin{pmatrix}v_k^* \\ u_k^* \end{pmatrix}||^2
\end{equation}
where $||\ldots||$ is the usual ${\cal L}_2$ norm and
$k_s$ is the index of the scissors mode.
The basic assumption is that the scissors mode
is the only one with vanishing frequency in the $\epsilon\rightarrow 0$
limit. We rewrite Eq.(\ref{eq:rule1}) as
\begin{equation}
\sum_{k\in {\cal F}_+} 2\epsilon_k |b_k|^2 =
2m\omega^2\epsilon \int (x^2-y^2) |\psi|^2.
\end{equation}
In the degenerate case the right-hand side is $O(\epsilon^2)$
so that each (positive) term $\epsilon_k |b_k|^2$ is $O(\epsilon^2)$.
For the normal modes, both $\epsilon_k^{-1}$ and the mode functions
$u_k,v_k$ have a finite limit for $\epsilon\rightarrow 0$, which 
proves $||\vec{e}_0^{\,\rm non-scissors}||^2=O(\epsilon^2)$.

In the non-degenerate case the same reasoning leads to a
weight being $O(\epsilon)$. A better estimate can be obtained
from
\begin{equation}
\langle\vec{e}_0,\vec{e}_1\rangle-\epsilon_{k_s}^2
\langle\vec{e}_0,\vec{e}_{-1}\rangle=
\sum_{k\in{\cal F}_+\setminus \{k_s\}} 2\left(\epsilon_k-\epsilon_{k_s}^2/\epsilon_k\right)
|b_k|^2.
\label{eq:rule_combi}
\end{equation}
Then using the explicit expression for the scalar products in the left-hand
side, see Eq.(\ref{eq:rule-1}) and Eq.(\ref{eq:rule1}),
and using the result Eq.(\ref{eq:nusquared}) of the perturbative
expansion for $\nu_{\rm scissors}=\epsilon_{k_s}/\hbar$,
one realizes that the left-hand side of Eq.(\ref{eq:rule_combi})
is $O(\epsilon^2)$, which leads to a weight on non-scissors modes
being $O(\epsilon^2)$ as in the degenerate case.


\begin{thebibliography}{99}
\bibitem{quasi2D} Quasi 2D means that the motion of the
gas along $z$ is frozen in the 
ground state of a trapping potential, but still the $s$-wave scattering length
$a$ remains much smaller than the spatial width of the gas along $z$
so that the scattering of two atoms does not acquire a true 2D character.
\bibitem{note_pour_carlos}
This shows that the zero frequency mode predicted by classical hydrodynamics
is not the scissors mode of the classical hydrodynamics.
Anticipating the quantum treatment of the problem, one can show that
for $\epsilon=0$, the zero frequency mode of the classical hydrodynamics actually 
corresponds to the anomalous mode $\vec{e}_a$ of the quantum theory.
\bibitem{tech2} The argument of the logarithm is not calculated in \cite{Aftalion1} but in 
\cite{Aftalion2} where it is done for the 3D case. Converting this calculation to 2D, we 
find that the argument does not depend on $\epsilon$ to first order.
\bibitem{o_vs_om} 
The condition that $|x_i|, |y_i|$ are much smaller than the Thomas-Fermi radius,
combined with the sum rule \cite{Aftalion1}
$\sum_i (1-\epsilon)x_i^2+(1+\epsilon) y_i^2=n_v(n_v-1)\lambda^2$
for the steady state positions,
implies that $(n_v-1)^{1/2}\lambda$ is smaller than the Thomas-Fermi radius,
which leads to the condition 
$(\Omega-\Omega_m)/\Omega_m \gg (n_v-1)/|\log\eta|$.
\bibitem{really_technical}
If one keeps to first order the position dependence of the prefactor
of the $\log d_{ij}$ in \cite{Aftalion1}, one gets an extra contribution
on the right hand side of Eq.(\ref{eq:x2my2_asympt}), given by $2\pi\eta^2 \sum_{i\neq j} (x_i^2-y_i^2)
\log(d_{ij}/l)$. We have checked for $n_v=2, 9$ and $10$ that
it is indeed negligible under the condition derived in \cite{o_vs_om}.
\bibitem{tech} The procedure is as follows: we wish to find $\vec{x}$ such
that $\vec{x}={\cal L}^{-1} \vec{s}$. It then follows that we need only
minimize the quantity $\vec{x}^* \cdot \eta {\cal L} \vec{x}-
(\vec{x}^* \cdot \eta \vec{s}+{\rm c.c.})$ where $\eta=\begin{pmatrix} 1 & 0
\\ 0 & -1 \end{pmatrix}$ since $\eta {\cal L}$ is a positive definite matrix
(because $\psi$ is a local minimum of the Gross-Pitaevskii energy functional)
whereas ${\cal L}$ is not. We then perform the minimization numerically 
using the conjugate gradient method.
\end{thebibliography}
\end{document}